\documentclass[aps,prd,showkeys,showpacs,amssymb,cite,
amsfonts,epsf,preprintnumbers,nofootinbib,superscriptaddress]{revtex4}

\usepackage[dvips]{graphicx}
\usepackage{bm,latexsym,amsmath,amssymb,amsfonts,color}

\newcommand{\be}{\begin{equation}}
\newcommand{\ee}{\end{equation}}
\newcommand{\bear}{\begin{eqnarray}}
\newcommand{\eear}{\end{eqnarray}}
\newcommand{\ba}{\begin{array}}
\newcommand{\ea}{\end{array}}
\newcommand{\nn}{\nonumber}

\begin{document}

\title{Tensor-to-Scalar Ratio
in Eddington-inspired Born-Infeld Inflation}

\author{Inyong Cho}
\email{iycho@seoultech.ac.kr}
\affiliation{Institute of Convergence Fundamental Studies \& School of Liberal Arts,
Seoul National University of Science and Technology, Seoul 139-743, Korea}
\author{Naveen K. Singh}
\email{naveen.nkumars@gmail.com}
\affiliation{Institute of Convergence Fundamental Studies \& School of Liberal Arts,
Seoul National University of Science and Technology, Seoul 139-743, Korea}

\begin{abstract}
We investigate the scalar perturbation of the inflation model
driven by a massive-scalar field in Eddington-inspired Born-Infeld gravity.
We focus on the perturbation at the attractor stage in which the first and the second slow-roll
conditions are satisfied.
The scalar perturbation exhibits the corrections to the chaotic inflation model
in general relativity.
We find that the tensor-to-scalar ratio becomes smaller than that of the usual chaotic inflation.
\end{abstract}
\maketitle

\section{Introduction}
The Eddington-inspired Born-Infeld (EiBI) gravity was recently developed in Ref.~\cite{Banados:2010ix}.
The action in this theory is described by
\begin{eqnarray}\label{action}
S_{{\rm EiBI}}=\frac{1}{\kappa}\int
d^4x\Big[~\sqrt{-|g_{\mu\nu}+\kappa
R_{\mu\nu}(\Gamma)|}-\lambda\sqrt{-|g_{\mu\nu}|}~\Big]+S_{\rm M}(g,\varphi),
\end{eqnarray}
where $\kappa$ is the only additional parameter of the theory,
and $\lambda$  is a dimensionless parameter related with the
cosmological constant by $\Lambda = (\lambda -1)/\kappa$.
This theory is based on the Palatini formalism in which
the metric $g_{\mu\nu}$ and the connection
$\Gamma_{\mu\nu}^{\rho}$ are treated as independent fields.
The Ricci tensor $R_{\mu\nu}(\Gamma)$ is evaluated
solely by the connection, and the matter field is coupled
only to the gravitational field $g_{\mu\nu}$.

The merit of EiBI gravity is that it is equivalent to
general relativity (GR) in the vacuum \cite{Banados:2010ix}.
Therefore, the astronomical phenomenon such
as the light deflection by a star is not altered.
More interesting cosmological consequence appears when it is applied to the Universe filled with perfect fluid.
It predicts a singularity-free initial state with a finite size
\cite{Banados:2010ix,Cho:2012vg}.

By introducing a massive scalar field to the Universe in this theory,
the inflationary feature was investigated in Ref.~\cite{Cho:2013pea}.
The matter action is in the usual form used for the chaotic inflation model \cite{Linde:1983gd}
in GR,
\be \label{S:chaotic}
S_{\rm M}(g,\varphi) = \int d^4 x \sqrt{-|g_{\mu\nu}|}
\left[ -\frac12 g_{\mu\nu} \partial^\mu\varphi \partial^\nu \varphi -V(\varphi) \right],
\qquad
V(\varphi) = \frac{m^2}{2} \varphi^2.
\ee
%
In EiBI gravity, there exists an upper bound in pressure
due to the square-root type of the action.
When the energy density is high,
the maximal pressure state (MPS) is achieved,
for which the scale factor exhibits an exponential expansion.
It was investigated in Ref.~\cite{Cho:2013pea} that this MPS is the past attractor
from which all the classical evolution paths of the Universe originate.
The energy density is very high in the MPS, but
the curvature scale remains constant since the Hubble parameter becomes $H_{\rm MPS} \approx 2m/3$.
Therefore, quantum gravity is not necessary in describing the high-energy state
of the early universe.

The MPS is unstable under the global perturbation (zero-mode scalar perturbation)
and evolves to an inflationary attractor stage.
The succeeding inflation feature is the same with the ordinary chaotic inflation in GR,
but it is not chaotic at the high-energy state
because the pre-inflationary stage can have a finite low curvature.
Depending on the initial conditions, the evolution of the Universe
can acquire the 60 $e$-foldings in the late-time inflationary attractor period.
If the sufficient  $e$-foldings are not acquired in this period,
it must be complemented in the exponentially expanding period at the near-MPS
in order to solve the cosmological problems.
%

Since this scalar-field model in EiBI gravity provides the natural and regular pre-inflationary
stage with the usual attractor inflationary stage,
it is worthwhile to investigate the density perturbation in order to
examine its consistency with the observational results.
The tensor perturbation in this model was investigated in Ref.~\cite{Cho:2014ija}.
For short wave-length modes, the perturbation is very similar to that of the usual
chaotic inflation in GR, with a small EiBI correction.
For long wave-length modes, however, there is a peculiar peak in the power spectrum
originated from the near-MPS stage.
This may leave a signature in the cosmic microwave background radiation.

In this paper, we investigate the scalar perturbation of this model.
(The density perturbation has been studied in the EiBI universe filled with perfect fluid
in Refs.~\cite{Lagos:2013aua,EscamillaRivera:2012vz,Avelino:2012ue,Yang:2013hsa}.
Other works have been investigated in the cosmological and
astrophysical aspects in
Refs.~\cite{Pani:2011mg,Pani:2012qb,DeFelice:2012hq,Avelino:2012ge,Avelino:2012qe,Casanellas:2011kf,
Liu:2012rc,Delsate:2012ky,Pani:2012qd,Cho:2013usa,Scargill:2012kg,Kim:2013nna,Kim:2013noa,Du:2014jka,Ji:2014hna}.)
From the very recent observational result of BICEP2 \cite{Ade:2014xna}
there is an increasing interest in the tensor-to-scalar ratio
of the various inflationary models.
Therefore, we focus on the scalar power spectrum at the inflationary ``attractor stage"
at which the main band for the test of the tensor-to-scalar ratio is produced.
We shall obtain the EiBI correction to the scalar power spectrum of the usual chaotic inflation.
With the result of the tensor perturbation obtained in Ref.~\cite{Cho:2014ija},
we get the tensor-to-scalar ratio in the EiBI inflationary model.

\section{Field equations}
The EiBI theory can be formulated as a bimetric-like theory
with the action,
\be\label{S}
S[g,q,\varphi] = \frac{1}{2} \int d^4x \sqrt{-|q_{\mu\nu}|} \left[  R(q) - \frac{2}{\kappa}  \right]
+{1 \over 2\kappa} \int d^4x \left(\sqrt{-|q_{\mu\nu}|}q^{\alpha\beta}g_{\alpha\beta} - 2\sqrt{-|g_{\mu\nu}|}
\right)+ S_{\rm M}[g,\varphi],
\ee
where $g_{\mu\nu}$ is the metric and $q_{\mu\nu}$ is the auxiliary metric.

When there is no cosmological constant ($\lambda =1$),
the action \eqref{action} is completely equivalent to the action \eqref{S}
if one considers that $\Gamma$ is the affine connection of $q_{\mu\nu}$.
The equations of motion are
\begin{align}
\frac{\sqrt{-|q|}}{\sqrt{-|g|}}~q^{\mu\nu}
& =\lambda g^{\mu\nu} -\kappa T^{\mu\nu},\label{eom1}\\
q_{\mu\nu} & = g_{\mu\nu}+\kappa R_{\mu\nu}, \label{eom2}
\end{align}
where $T^{\mu\nu}$ is the standard energy-momentum tensor.
The ans\"atze for the auxiliary metric and the metric are
\begin{align}
q_{\mu\nu}dx^\mu dx^\nu &= b^2(\eta) \left[ -\frac{d\eta^2}{z(\eta)} +\delta_{ij}dx^idx^j \right],\\
g_{\mu\nu}dx^\mu dx^\nu
&= -dt^2 +a^2(t) \delta_{ij}dx^idx^j
= a^2(\eta) \left( -d\eta^2 +\delta_{ij}dx^idx^j \right),
\end{align}
where $t$ is the cosmological time and $\eta$ is the conformal time for the metric.
The derivatives are defined as $\hat{} \equiv d/dt$ and $' \equiv \d/d\eta$.
In this paper, we shall denote ${\cal H} \equiv a'/a$, $H \equiv \hat{a}/a$,
$h\equiv b'/b$, and $h_b \equiv \hat{b}/b$.
The components of Eq.~\eqref{eom1} give
\be\label{eom1cp}
b^2\sqrt{z}=\left(1+ \kappa\rho_0\right)a^2,
\qquad
\frac{b^2 }{\sqrt{z}}=\left(1 - \kappa p_0\right)a^2,
\ee
where we denote the subscript $0$ for the unperturbed background scalar field, so
$\rho_0 = \varphi_0'^2/2a^2 + V(\varphi_0)$ and $p_0 = \varphi_0'^2/2a^2 - V(\varphi_0)$.
From Eq.~\eqref{eom1cp}, one gets $z=(1+\kappa\rho_0)/(1 - \kappa p_0)$.
The components of Eq.~\eqref{eom2} provide dynamical equations,
\begin{align}
b^2 &= 3\kappa z \left(\frac{b'}{b}\right)^2 +\frac{a^2}{2} (3-z), \label{eqn_b1}\\
b^2 &= a^2 +\kappa z \left[ \frac{b''}{b} + \left(\frac{b'}{b}\right)^2
+ \frac{1}{2}\frac{b'}{b}\frac{z'}{z}\right], \label{eqn_2}
\end{align}
and the scalar field equation is given by
\be\label{Seq}
\varphi_0''+2{\cal H}\varphi_0' +a^2 \frac{dV}{d\varphi_0} =0.
\ee
The background fields, $a$, $b$, $z$, and $\varphi_0$ are obtained by solving
Eqs. \eqref{eom1cp}-\eqref{Seq}.

Now let us consider the scalar perturbation.
The perturbation fields for $q_{\mu\nu}$ and $g_{\mu\nu}$ are defined as
\begin{align}
ds_q^2 &= b^2\left\{-\frac{1+ 2\phi_1}{z} d\eta^2  +2\frac{B_{1,i}}{\sqrt{z}} d\eta dx^i
+ \Big[(1-2\psi_1)\delta_{ij} + 2 E_{1,ij}\Big] dx^i dx^j \right\}, \label{qpert} \\
ds_g^2 &= a^2\left\{-(1+ 2\phi_2) d\eta^2 + 2 B_{2,i}d\eta dx^i
+\Big[(1-2\psi_2)\delta_{ij} + 2 E_{2,ij}\Big]dx^i dx^j \right\},  \label{metric_pert}
\end{align}
and the perturbation for the scalar field is given by $\varphi=\varphi_0 +\chi$.
Therefore, there are nine perturbation fields in total.
Let us denote them as $F_l$, where $l=1\sim 9$.
With the perturbed metrics and the scalar field,
one can expand the action \eqref{S} up to the second order in the perturbation fields.
Then the second-order action can be collected as $S_{\rm s}=S_1+S_2+S_3$ where
$S_1$ contains the perturbation fields for $q_{\mu\nu}$,
$S_2$ contains the perturbation fields for $g_{\mu\nu}$
and the mixing terms for $q_{\mu\nu}$,
and $S_3$ contains the matter field perturbation (see Ref.~\cite{Lagos:2013aua}),

\begin{align}\label{S1}
S_{1}[\phi_1,B_1,\psi_1,E_1]
&= \frac{1}{2}\int d^4x\; \Bigg\{  \frac{b^2}{\sqrt{z}} \left[4zh\psi_1' E_{1,ii}-6z\psi_1^{'2} \right.
-12zh(\phi_1+\psi_1)\psi'_1-2\psi_{1,i}(2\phi_{1,i}-\psi_{1,i} ) \nonumber \\
&- 4h\psi_{1,i}B_{1,i} +6zh^2(\phi_1+\psi_1)E_{1,ii}-4\sqrt{z}h(\phi_1+\psi_1)(B_1-\sqrt{z}E_1')_{,ii} \nonumber \\
&- 4\sqrt{z}\psi_1'(B_1-\sqrt{z}E_1')_{,ii}
-4\sqrt{z}h E_{1,ii}(B_1-\sqrt{z} E'_1)_{,jj}
 +4\sqrt{z}hE_{1,ii}B_{1,jj} \nonumber \\
&+ 3zh^2E_{1,ii}E_{1,jj} + 3zh^2B_{1,i}B_{1,i} \left.-9zh^2(\phi_1+\psi_1)^2\frac{}{}\right] \nonumber \\
&- \frac{2 b^4}{\kappa\sqrt{z}}\left[\frac{3}{2}\psi_1^2-3\phi_1\psi_1+\frac{1}{2}B_{1,i}B_{1,i}\right.
 \left.-\frac{1}{2}E_{1,ii}E_{1,jj}-\frac{1}{2}\phi_1^2+E_{1,ii}(\phi_1-\psi_1)\right] \Bigg\},
\end{align}
\begin{align}\label{S3}
S_{2}[\phi_k,B_k,\psi_k,E_k]
&= \frac{1}{2}\int d^4x \; \Bigg\{ \frac{a^2 b^2}{\kappa\sqrt{z}} \Big[2\sqrt{z}B_{1,i}B_{2,i}
+ \phi_1\left[\left(z-1\right)\left(3\psi_1-E_{1,ii}\right)-6\psi_2  +2E_{2,ii}-2z\phi_2\right] \nonumber \\
&+ \psi_1 \left[6\psi_2-(z-1)E_{1,ii}-2E_{2,ii}-6z\phi_2\right] - \frac{1}{2}(z-1)(E_{1,ii}E_{1,jj}+B_{1,i}B_{1,i})  \nonumber \\
&+ \frac{3}{2}\left(\phi_1^2+\psi_1^2\right)(z-1) -  2E_{1,ii}\left(\psi_2-z\phi_2+E_{2,ii}\right)\Big]
 \nonumber \\
&-\frac{2a^4}{\kappa}\left[\frac{3}{2}\psi^2_2-\frac{1}{2}\phi_2^2
+ \frac{1}{2}B_{2,i}B_{2,i} -\frac{1}{2}E_{2,ii}E_{2,jj}+(\phi_2-\psi_2)E_{2,ii} -3\phi_2\psi_2\right] \Bigg\},
\end{align}
\begin{align}\label{S3}
S_{3}[\phi_2,B_2,\psi_2,E_2,\chi]
&= \frac{1}{2} \int d^4x\; a^2\Bigg\{ \varphi_0'^2 \left(4 \phi_2^2 - B_{2,i}B_{2,i} \right) \nonumber \\
&+ \left(\varphi_0'^2 - 2 V_0 a^2\right) \left[ \frac{1}{2} \left( 3 \psi_2^2 - \phi_2^2
+ B_{2, i} B_{2, i}- E_{2,ii}E_{2,ii}\right) - 3 \phi_2 \psi_2 + (\phi_2-\psi_2) E_{2,ii}\right]\nonumber \\
&-2 \varphi_0' \chi_{, i} B_{2,i} - 4 \varphi_0' \chi' \phi_2 + \chi'^2
+ 2 \left(\phi_2-3\psi_2 + E_{2, ii}\right) (\chi' \varphi_0' - V_1 a^2 - \phi_2 \varphi_0'^2) -\chi_{, i}\chi_{, i} - 2V_2 a^2 \Bigg\},
\end{align}
where $V_i$ is the $i$th-order potential
from $V=V_0(\varphi_0)+V_1(\chi)+V_2(\chi)$.
Using $\rho_0$ and $p_0$, $S_3$ can be recast into
\begin{align}
S_{3}[\phi_2,B_2,\psi_2,E_2,\chi]
&= \int d^4 x a^4 \Bigg\{
p_0 \left[ \frac{1}{2} \left( 3 \psi_2^2 - \phi_2^2+ B_{2, i} B_{2, i}- E_{2,ii}E_{2,ii}\right)
- 3 \phi_2 \psi_2 + (\phi_2-\psi_2) E_{2,ii}\right]  \nonumber \\
&+ (\rho_0 +p_0) \left[ 2\phi_2(\phi_2 -{\cal X}\chi') -\frac{1}{2}B_{2,i}(B_{2,i} +2{\cal X}\chi_{,i})
+(\phi_2-3\psi_2 + E_{2, ii})({\cal X}\chi' -{\cal Y}\chi - \phi_2)   \right] \nonumber \\
&+ \frac{1}{2 a^2}(\chi'^2-\chi_{,i} \chi_{,i}) - \frac{m^2}{2} \chi^2\Bigg\},
\end{align}
where ${\cal X} \equiv 1/(a\sqrt{\rho_0+p_0})$ and ${\cal Y} \equiv- m\sqrt{\rho_0-p_0}/(\rho_0+p_0)$.

For the nine perturbation fields, $F_l$, we introduce the corresponding Fourier modes as
\be\label{Fourier}
F_l(\eta,\vec{x}) = \int \frac{d^3k}{(2\pi)^{3/2}}
F_l(\eta,\vec{k})e^{i\vec{k}\cdot\vec{x}}.
\ee
There is the gauge freedom in the action for the perturbation fields.
The gauge conditions have been studied precisely in Ref.~\cite{Lagos:2013aua}.
Since we consider a scalar field, the proper gauge choice from the study of Ref.~\cite{Lagos:2013aua} 
is to fix the value of one element from each set given as below,
\bear
\left(\psi_1, \psi_2, \chi\right) + \left(E_1, E_2\right).
\eear
In this work we choose the gauge conditions as
\be\label{gg}
\psi_1=0 \quad{\rm and}\quad E_1=0,
\ee
which provide a decoupled field equation in the end.
Then from the variation of $S_2$ and $S_3$ for $\phi_2$, $\psi_2$, $E_2$, and $B_2$,
we get
\begin{align}
& (1-2z)\phi_2 +z\phi_1 -3z\psi_2 -k^2zE_2 -(1-z){\cal X}\chi' -(1-z){\cal Y}\chi =0, \label{eqn_phi2} \\
& 3\psi_2 + 3\phi_1 - 3z\phi_2 +k^2E_2 - 3(1-z){\cal X}\chi' +3(1-z){\cal Y}\chi =0, \label{eqn_psi2}\\
& k^2E_2 - \phi_1 + z\phi_2 - \psi_2 +(1-z){\cal X}\chi' -(1-z){\cal Y}\chi =0, \label{eqn_E2}\\
& zB_2 - \sqrt{z}B_1 -(1-z){\cal X}\chi =0. \label{eqn_B2}
\end{align}
From the variation of $S_1$ and $S_2$ for $\phi_1$ and $B_1$,
we get
\begin{align}
& (6\kappa h^2 -a^2)z\phi_1 +a^2z\phi_2 +3a^2\psi_2 +k^2a^2E_2 -2k^2\kappa h\sqrt{z}B_1 =0, \label{eqn_phi1} \\
& a^2B_1 -2\kappa h\sqrt{z}\phi_1 - a^2\sqrt{z}B_2  =0. \label{eqn_B1}
\end{align}
From Eqs. \eqref{eqn_psi2} and \eqref{eqn_E2}, we have $E_2 = 0$.
From Eqs. \eqref{eqn_phi2} and \eqref{eqn_E2}, we have
\be
\phi_2 = \frac{(z-1)(3z+1){\cal X}\chi' -(z-1)(3z-1){\cal Y}\chi +4z\phi_1}{(z+1)(3z-1)} \label{int_phi2},
\ee
and from Eqs. \eqref{eqn_B2} and \eqref{eqn_B1}, we have
\be
\phi_1 = \frac{a^2(z-1){\cal X}\chi}{2 \kappa hz}.
\ee
Then from Eqs. \eqref{eqn_psi2} and \eqref{int_phi2}, we finally get
\be\label{psi2_XY}
\psi_2 = \frac{z-1}{2\kappa hz(z+1)(3z-1)}
\Big[ -2\kappa hz(z-1){\cal X}\chi' +a^2(z-1)^2{\cal X}\chi +2\kappa hz(3z-1){\cal Y}\chi \Big],
\ee
which is expressed only by the background fields and the matter-field perturbation $\chi$.
This quantity will be used later in evaluating the power spectrum from the comoving curvature,
\be
{\cal R} = \psi_2 + \frac{H}{\hat{\varphi}_0}\chi.
\ee

With the results of Eqs.~\eqref{eqn_phi2}-\eqref{psi2_XY},
we can write the second-order action $S_{\rm s}[\chi]$ expressed only
by the matter-field perturbation $\chi$ and the background fields in the Fourier space,
\be\label{Ss}
S_{\rm s}[\chi] = \frac{1}{2}\int  d^3k d\eta\;   \Big[ f_1(\eta,k) \chi'^{2} - f_2(\eta,k)  \chi^2 \Big],
\ee
where
\be\label{f1}
f_1(\eta,k) = a^2 +\frac{2a^2(z-1)^2{\cal X}^2\left[ a^2(z-3)-6\kappa h^2z \right]}{\kappa\sqrt{z}(z+1)(3z-1)},
\ee
and
\be
f_2(\eta,k) = \frac {\beta}{8\kappa^3 h^2 z^{5/2}(z + 1)^2}.
\ee
Here,
\be
\beta= a^2 \left[\frac{\beta_1}{3z-1} + \frac{\beta_2}{(3z-1)^2}\right],
\ee
where
\begin{align}
\beta_1 &= (z+1)\Bigg\{ 8\kappa^3h^2z^2(3z-1)\Big[ k^2\sqrt{z} - 12 h^2{\cal Y}^2 z
+ k^2 z^{3/2} + 24 h^2 {\cal Y}^2 z^2 - 12 h^2 {\cal Y}^2 z^3 - 3 k^2 h^2  {\cal X}^2 (z-1)^2(z+1)\Big]  \nonumber \\
&+ a^6 {\cal X}^2 (z-3)(z-1)^3(3z^2-2z +3)
+ 4 \kappa a^4 h {\cal X} z(z-1)^2 \Big[ {\cal Y}(z-3)^2(3z-1) - 3 h {\cal X} z(3z^2 -6z -1)\Big] \nonumber \\
 &+ 4 \kappa^2 a^2 h^2 z(3z-1) \Big[-6h {\cal X}{\cal Y}(z-3)(z-1)^2 z + {\cal X}^2(z-1)^2(z+1)[(k^2+ 9h^2)z -3k^2] \nonumber \\
 &+ 4{\cal Y}^2z(z-3)(z-1)^2   + 2\kappa m^2 z^{3/2}(z+1) \Big]\Bigg\}, \\
\beta_2 &= (z-1) \Big[a^2 ( z-3) - 6 \kappa h^2  z\Big]
\Bigg\{ a^4 {\cal X}^2 (z-1)^2(z+1)(3z-1)(3z^2-2z+3)\nonumber \\
&+ 4 \kappa^2 h^2 z(3z-1)^2 \Big[2z(z-1)(z+1) [ 2(h+\mathcal{H}){\cal X}{\cal Y} +\left({\cal X}{\cal Y}\right)']
+ {\cal X}{\cal Y}(z^2 + 6z + 1) z' \Big] \nonumber \\
&+ 2 \kappa a^2 h {\cal X} \Big[z(z-1)(z+1)(3z-1)(3z^2-2z+3) [(h + 4{\cal H}){\cal X} + 2{\cal X}']  \nonumber \\
&+ {\cal X}(9 z^5 + 21z^4 - 34 z^3 + 30 z^2 + 9 z  -3)z'\Big] \Bigg\}.
\end{align}

The field $\chi$ in the action \eqref{Ss} is not of the canonical form.
Therefore, we introduce the canonical field $Q$ by the transformation $\chi = Q/\omega$
with introducing a new time coordinate $\tau$ by $d\eta =f_3d\tau$.
Then the field equation becomes
\be\label{EOM1}
\ddot{Q} + \left( \frac{\dot{f_1}}{f_1}-\frac{\dot{f_3}}{f_3} -2\frac{\dot{\omega}}{\omega}\right)\dot{Q}
+ \left[ \frac{f_2 f_3^2}{f_1} -\frac{\dot\omega}{\omega}
\left( \frac{\dot{f}_1}{f_1} -\frac{\dot{f}_3}{f_3} -2\frac{\dot\omega}{\omega} \right)
-\frac{\ddot\omega}{\omega}  \right]Q =0.
\ee
For the canonical field, the $\dot Q$-term vanishes, and thus
we get $\omega^2 =f_1/f_3$.
The field equation then becomes
\be\label{Q-eq1}
\ddot{Q} + \left(\frac{f_1 f_2}{\omega^4}-\frac{\ddot{\omega}}{\omega}\right)Q
\equiv \ddot{Q} + \left(\sigma_s^2 k^2-\frac{\ddot{\omega}}{\omega}\right)Q =0 ,
\ee
where $\sigma_s^2 \equiv f_1 f_2/k^2\omega^4$.
We assume a Bunch-Davies vacuum described by the plane wave,
requiring $\sigma_s^2 \to 1$ in the limit of $k\to\infty$.
Then $\omega$ is determined as
\begin{align}\label{omega}
\omega^4 &= \frac{a^4}{2 \kappa^2 z^2(z+1)(3z-1)} \Bigg\{ a^2 {\cal X}^2 (z-3)(z-1)^2
- 2 \kappa z \Big[ 3 h^2 {\cal X}^2 (z-1)^2 -\sqrt{z}\Big]\Bigg\} \nonumber \\
 &\times \Bigg\{ 2 a^2 {\cal X}^2 (z-3)(z-1)^2
 - \kappa \sqrt{z}\Big[12 h^2 {\cal X}^2 \sqrt{z} (z-1)^2  -3 z^2 -2 z +1 \Big]\Bigg\}.
\end{align}
For the canonical field $Q$, the normalization condition is given by
\be\label{norm}
Q\dot Q^*-Q^*\dot Q =i.
\ee
For the initial perturbation produced in the Bunch-Davies vacuum ($k\to\infty$ and $\sigma_s^2 \to 1$),
we impose the minimum-energy condition
which picks up the positive mode solution of Eq.~\eqref{Q-eq1}.
The normalization condition \eqref{norm} fixes the coefficient,
and the solution becomes
\be\label{solBD}
Q(\tau) = \frac{e^{-ik\tau}}{\sqrt{2k}}.
\ee

\section{Perturbation at Attractor Stage}
At the attractor stage,
both of the first and the second slow-roll conditions are satisfied.
The background evolution was found \cite{Cho:2013pea} to be approximately
the same with that of the usual chaotic inflation in GR.
In this paper, we focus on the scalar perturbation at the attractor stage,
and investigate the EiBI correction in the power spectrum.

At the attractor stage,
the background scalar field and scale factor are given by
\be\label{slowsol}
\varphi_0(t) \approx \varphi_i +\sqrt{\frac{2}{3}}m t,
\qquad
a(t) \approx a_i\; e^{[\varphi_i^2-\varphi_0^2(t)]/4},
\ee
where $\varphi_i <0$ is the value of the scalar field in the beginning of the attractor.
(We consider the scalar field rolling down the potential at $\varphi <0$.)
For 60 $e$-foldings, $|\varphi_i| \gtrsim 15$ is required.
From observational data, $m \sim 10^{-5}$ for the standard inflationary model.
At the early stage of the attractor, $m^2t^2 \ll mt$.
(We set $t=0$ as the beginning of the attractor stage.)
Then the scale factor is further approximated as
\be\label{aATT}
a(t) \approx a_i e^{-\varphi_i mt/\sqrt{6} - m^2t^2/6}
\approx a_i e^{-\varphi_i mt/\sqrt{6}}.
\ee
In this background, we can approximate $z$ as
\be\label{z_approx}
z= \frac{1+\kappa\rho_0}{1 - \kappa p_0}
= \frac{1+\kappa (\hat{\varphi}_0^2/2 + m^2\varphi_0^2/2)}{1-\kappa (\hat{\varphi}_0^2/2 - m^2\varphi_0^2/2)}
= 1 + \frac{\kappa\hat{\varphi}_0^2}{1-\kappa (\hat{\varphi}_0^2/2 - m^2\varphi_0^2/2)}
\approx 1 + \frac{2\kappa m^2/3}{1+\kappa m^2\varphi_0^2/2},
\ee
where we used the first slow-roll condition $\hat{\varphi}_0^2/2 \ll m^2\varphi_0^2/2$ in the last step.
At the attractor stage, therefore, the value of $z-1$ is a small quantity proportional to $\kappa$,
which is responsible for the ``EiBI correction" in the power spectrum
as we will see in the next section.

Now, let us evaluate the involved quantities in approximation.
From the background field equations in Eq.~\eqref{eom1cp},
we have
\be\label{b_approx}
b = (1+\kappa\rho_0)^{1/4}(1-\kappa p_0)^{1/4}a
\approx (1+\kappa\rho_0)^{1/2}a
\approx a\sqrt{1+\frac{1}{2}\kappa m^2\varphi_0^2},
\ee
where we used the first slow-roll condition for the approximation.
Then the scalar factor $h_b$ can be approximated by
\be
h_b = \frac{\hat{b}}{b}
= \frac{\hat{a}}{a} +\frac{\kappa m^2\varphi_0\hat{\varphi}_0}{2(1+\kappa m^2\varphi_0^2/2)}
\approx -\frac{m\varphi_i}{\sqrt{6}} +\frac{\kappa m^3\varphi_i}{\sqrt{6}}
\approx -\frac{m\varphi_i}{\sqrt{6}} = \frac{\hat{a}}{a} =H,
\ee
where we assumed $\kappa m^2 \ll 1$.
Therefore, the terms containing $\kappa h_b^2 \approx \kappa m^2\varphi_i^2/6 \ll 1$
can be ignored in the approximations.
Using Eqs. \eqref{eom1cp} and \eqref{b_approx}, we get
\begin{align}
{\cal X}^2 &= \frac{1}{a^2(\rho_0+p_0)}=\frac{\kappa\sqrt{z}}{b^2(z-1)}
\approx \frac{\kappa\sqrt{z}}{a^2(z-1)}\left( 1+\frac{1}{2}\kappa m^2\varphi_0^2 \right)^{-1}
\approx \frac{\kappa\sqrt{z}}{a^2(z-1)}\left( 1-\frac{1}{2}\kappa m^2\varphi_i^2 \right),\label{X_approx}\\
{\cal Y} &=-m \frac{\sqrt{\rho_0-p_0}}{\rho_0+p_0}
\approx -m \left[ \frac{z+1}{\kappa\sqrt{z}}\left( 1+\frac{1}{2}\kappa m^2 \varphi_0^2 \right) -\frac{2}{\kappa} \right]^{1/2}
\left[ \frac{z-1}{\kappa\sqrt{z}}\left( 1+\frac{1}{2}\kappa m^2 \varphi_0^2 \right) \right]^{-1} \nn\\
&\approx- \frac{m}{(z-1)} \left[\kappa \sqrt{z}(z+1) \left( 1-\frac{1}{2}\kappa m^2 \varphi_i^2 \right)
-2\kappa z \left( 1-\kappa m^2 \varphi_i^2 \right) \right]^{1/2},
\label{Y_approx}
\end{align}
and $\omega$ in Eq.~\eqref{omega} can be approximated as
\be\label{omega_approx}
\omega \approx a S,
\quad{\rm where}\quad
S \equiv \Bigg[ \frac{(z^2-2z+3)(5z^2-6z+5)}{2z(z+1)(3z-1)}\Bigg]^{1/4}.
\ee
Plugging Eq.~\eqref{X_approx} into Eq.~\eqref{f1}
and reminding of $\kappa m^2 \ll1$ and $\kappa h^2 = \kappa h_b^2a^2 \ll a^2$,
we get
\be\label{f1_approx}
f_1 \approx a^2 \frac{5z^2-6z+5}{(z+1)(3z-1)}.
\ee
Then from Eqs.~\eqref{omega_approx} and \eqref{f1_approx}, we get
\be\label{f3_approx}
f_3 =\frac{f_1}{\omega^2} \approx \left[ \frac{2z(5z^2-6z+5)}{(z+1)(3z-1)(z^2-2z+3)} \right]^{1/2}.
\ee

Now let us keep the lowest-order correction that is proportional to $\kappa m^2$.
Then from Eqs.~\eqref{f1_approx} and \eqref{f3_approx}, we have
\be\label{f13omega}
f_1 \approx a^2 \left( 1-\frac{2}{3}\kappa m^2 \right),
\quad
f_3 \approx 1 + \frac{2}{9}(\kappa m^2)^2 \approx 1,
\quad\mbox{and}\quad
\omega^4 \approx a^4 \left( 1-\frac{4}{3}\kappa m^2 \right).
\ee
Using the results for ${\cal X}$ and ${\cal Y}$
in Eqs.~\eqref{X_approx} and \eqref{Y_approx},
we can get
\be\label{f2_approx}
f_2 \approx a^2\left\{ k^2\left( 1-\frac{2}{3}\kappa m^2 \right)
- m^2a^2 \Big[ 1+2\kappa m^2 (\varphi_i^2-1) \Big] \right\}.
\ee
For the time transformation at the attractor stage, we have then
\be
d\eta =f_3d\tau \approx d\tau.
\ee

\section{Power Spectrum}
In this section, we evaluate the scalar power spectrum at the attractor stage
using the quantities that we obtained in the previous section.
We shall focus on the corrections from the EiBI theory,
and compare the result with the power spectrum in GR.
Finally we will get the EiBI correction in the tensor-to-scalar ratio.

Let us express the scale factor $a$ in terms of $\tau$.
Using $a(t)$ in Eq.~\eqref{slowsol},
the time coordinates are transformed by
\be\label{tau-t}
d\tau = \frac{d\eta}{f_3} = \frac{dt}{f_3a} \approx \frac{dt}{a}
\quad\Rightarrow\quad
\int_{\tau_i}^\tau d\tau' = \int_0^t \frac{dt'}{a(t')}
\quad\Rightarrow\quad
\tau-\tau_i = \frac{\sqrt{6}}{\varphi_im} \left( \frac{1}{a} -\frac{1}{a_i} \right),
\ee
where we assumed that the attractor stage begins at $\tau=\tau_i>0$ ($t=0$).
[We assume that the Universe begins at the near-MPS stage at $\tau =0$ ($t\to -\infty$).]
Setting $t=0$ for the beginning of the attractor stage fixes
the arbitrariness of the scale factor, $a(t=0)=a_i$.
From Eq.~\eqref{tau-t}, the scale factor can be obtained as
\be\label{aATT}
a(\tau) = \frac{a_i(\tau_i-\tau_0)}{\tau-\tau_0},
\qquad
\tau_0 \equiv \tau(t\to\infty) = \tau_i - \frac{\sqrt{6}}{\varphi_im a_i}.
\ee

Let us consider the corrections for $\sigma_s^2$ and $\ddot\omega/\omega$,
in the field equation,
\be\label{Q-eq2}
\ddot{Q} + \Omega_k^2 Q =0,
\quad\mbox{where}\quad
\Omega_k^2 = \sigma_s^2 k^2-\frac{\ddot{\omega}}{\omega}.
\ee
For $\sigma_s^2$,
from the approximated quantities in Eqs.~\eqref{f13omega} and \eqref{f2_approx},
we get
\be\label{cs_approx}
\sigma_s^2 = 1-\frac{m^2a^2}{k^2}\left[ 1+2\kappa m^2\left( \varphi_i^2 -\frac{2}{3} \right) \right].
\ee
Here, the first term corresponds to the speed of sound, $c_s^2 =1$,
the $a^2$-dependence originates from the {\it non-conventional form} of the action,
and in particular, the $\kappa$-dependence is the EiBI correction.
Using the last expression for $z$ in Eq.~\eqref{z_approx} for $\omega$ in Eq.~\eqref{omega_approx}
and the time transformation between $t$ and $\tau$ in Eq.~\eqref{tau-t},
we get
\be\label{ddomega_approx}
\frac{\ddot\omega}{\omega}
\approx (1-\kappa^2m^4) \frac{\varphi_i^2 m^2a^2}{3}
\approx \frac{\varphi_i^2 m^2a^2}{3},
\ee
where we neglected the $\kappa$-dependent EiBI correction in the last step
since it is the higher-order in $\kappa$.
Therefore, the field equation \eqref{Q-eq2} can be approximated by
\begin{align}\label{Omega2_approx}
\Omega_k^2 &\approx  k^2
- \frac{\varphi_i^2 m^2a^2}{3} \left[ 1+ \frac{3}{\varphi_i^2}
+6\kappa m^2 \left(1-\frac{2}{3\varphi_i^2} \right) \right]
\approx k^2 - \frac{2}{(\tau-\tau_0)^2},
\end{align}
where we neglected the last three terms in the bracket
as $\varphi_i \sim {\cal O}(10)$ and $\kappa m^2 \ll 1$.
Therefore, there is no significant correction in the field equation,
and thus the normalized positive-energy mode solution
to the field equation \eqref{Q-eq2} becomes the usual one in GR,
\be\label{Q_approx}
Q \approx \frac{e^{-ik(\tau-\tau_0)}}{\sqrt{2k}} \left[ 1 - \frac{i}{k(\tau-\tau_0)}\right].
\ee

Now let us evaluate the comoving curvature.
Using Eqs.~\eqref{X_approx} and \eqref{Y_approx},
the most dominant term for $\psi_2$ is the last term in Eq.~\eqref{psi2_XY}.
Then we have
\be\label{psi2_approx}
\psi_2 \approx \frac{1}{2}\kappa m^2\varphi_i \chi
\qquad\Rightarrow\qquad
{\cal R} = \psi_2 + \frac{H}{\hat{\varphi}_0}\chi
\approx \frac{\kappa m^2-1}{2} \varphi_i \chi.
\ee
Here, $\psi_2$ results purely from the EiBI correction.
When $\kappa \to 0$, we have $\psi_2 \to 0$
which indicates that our choice of gauge condition ($\psi_1=0$ and $E_1=0$)
corresponds to the spatially flat gauge ($\psi_2=0$ and $E_2=0$) in the GR limit.

With the field $Q$ and the comoving curvature ${\cal R}$
obtained in Eqs. \eqref{Q_approx} and \eqref{psi2_approx},
the power spectrum is evaluated as
\begin{align}
P_{\cal R} &= \frac{k^3}{2 \pi^2} \mathcal{R}^2
= \frac{k^3}{2 \pi^2} \left(\psi_2  + \frac{H}{\hat{\varphi}_0} \chi\right)^2
\approx (1-\kappa m^2)^2\frac{k^3\varphi_i^2}{8\pi^2} \chi^2
= (1-\kappa m^2)^2\frac{k^3\varphi_i^2}{8\pi^2} \left( \frac{Q}{\omega} \right)^2\\
&\approx \frac{(1-\kappa m^2)^2}{(1-4\kappa m^2/3)^{1/2}} \frac{k^3\varphi_i^2}{8\pi^2} \left( \frac{Q}{a} \right)^2\\
&\approx \frac{(1-\kappa m^2)^2}{(1-4\kappa m^2/3)^{1/2}} \times  \frac{m^2\varphi_i^4}{96\pi^2}
\times  k^2(\tau-\tau_0)^2 \left[ 1+\frac{1}{k^2(\tau-\tau_0)^2} \right].
\end{align}
At the end of inflation ($\tau\to\tau_0$), finally we get
\begin{align}
P_{\cal R} = \frac{(1-\kappa m^2)^2}{(1-4\kappa m^2/3)^{1/2}} \times  \frac{m^2\varphi_i^4}{96\pi^2}
=\frac{(1-\kappa m^2)^2}{(1-4\kappa m^2/3)^{1/2}} \times P_{\cal R}^{\rm GR}
\approx \left(1-\frac{4}{3}\kappa m^2 \right) P_{\cal R}^{\rm GR},
\end{align}
where $P_{\cal R}^{\rm GR} = m^2\varphi_i^4/96\pi^2$ is the power spectrum in GR.

The tensor-to-scalar ratio is obtained with the result of the tensor power spectrum
obtained in Ref.~\cite{Cho:2014ija},
\be\label{r_EiBI}
r = \frac{P_{\rm T}}{P_{\cal R}}
\approx \frac{P_{\rm T}^{\rm GR}/(1+\kappa m^2\varphi_i^2/2)}{(1-4\kappa m^2/3)P_{\cal R}^{\rm GR}}
\approx \left( 1-\frac{1}{2}\kappa m^2\varphi_i^2 +\frac{4}{3}\kappa m^2 \right) r^{\rm GR},
\ee
where $r^{\rm GR} \sim 0.131$ for $60$ $e$-foldings.
The EiBI correction of the tensor spectrum lowers the value of $r$,
while that of the scalar spectrum raises the value.
As $\varphi_i \sim {\cal O}(10)$, the effect of the tensor spectrum is larger
and the whole EiBI corrections lower the value of $r$.

\section{Conclusions}
Recently the gravitational waves produced in the inflationary stage of the early Universe
attract much attention due to the observational result of BICEP2 \cite{Ade:2014xna}.
The result tells that the tensor-to-scalar ratio is very high, $r\sim 0.2$.
Although its validity requires further examinations, for example,
from the PLANCK observational results \cite{Ade:2013ktc},
it is very interesting to discuss how the various inflationary models predict the
value of the tensor-to-scalar ratio.

In this paper, we investigated the scalar perturbation in a newly suggested inflationary model
driven by a massive scalar field in Eddington-inspired Born-Infeld gravity \cite{Cho:2013pea}.
With the result of the tensor perturbation investigated in Ref.~\cite{Cho:2014ija},
we evaluated the tensor-to-scalar ratio.
As it was investigated in Ref.~\cite{Cho:2013pea},
there are two exponentially expanding stages of the Universe in this inflationary model.
The one is the near-MPS stage, and the other is the attractor stage.
We mainly focused on the attractor stage
since the main band for the test of the tensor-to-scalar ratio
is related with this stage.
(The near-MPS stage affects mostly the very long wave-length modes.)

The background evolution at the attractor stage
is very similar to that of the chaotic inflation in GR.
We assumed that the attractor stage maintained sufficiently long,
and investigated the scalar perturbation produced at this stage.
(For the perturbation produced at the near-MPS stage~\cite{Prep},
the result is very similar except for the very long wave-length modes
for which there exists a peculiar peak in the power spectrum
as in the tensor perturbation in Ref.~\cite{Cho:2014ija}.)
We assumed that the Bunch-Davies vacuum for the initial production
of the perturbation mode $k\to\infty$.
Then $\sigma_s$ was obtained accordingly for the arbitrary $k$-modes.
We imposed the minimum-energy condition for the intial perturbation
which picks up the positive-energy mode.

For the arbitrary $k$-modes, we obtained the EiBI corrections
in terms of $\kappa m^2$ which was assumed to be small.
(The strongest constraint for the value of $\kappa$ known so far, is from
the study of the star formation \cite{Avelino:2012ge,Pani:2011mg,Pani:2012qb},
$\kappa < 10^{-2} m^5 kg^{-1}s^{-2}$.
However, this is a very flexible constraint in Planck unit, $\kappa \lesssim10^{77}$.
As $m\sim 10^{-5}$ from observational data, the value of $\kappa m^2$ can have a wide range.)
The correction for the canonical perturbation field $Q$ is very tiny and minor,
so $Q$ is of the same form with that of the $\varphi^2$ chaotic inflation model in GR.
The main EiBI correction comes from two sources.
The one is from the relation $\chi = Q/\omega \equiv Q/aS$
between the matter field perturbation $\chi$ and its canonical form field $Q$.
In GR, $S=1$ while in EiBI $S\approx 1-\kappa m^2/3$.
The other is from the metric perturbation field $\psi_2$ in the comoving curvature
${\cal R} = \psi_2 +(H/\hat{\varphi}_0)\chi$.
This is related with the gauge.
In the spatially flat gauge in GR, $E_2=0$ and $\psi_2=0$.
In EiBI, we imposed the gauge conditions, $E_1=0$ and $\psi_1=0$,
which results in $E_2=0$ and $\psi_2=\kappa m^2\varphi_i\chi/2$.

With these corrections, the scalar power spectrum $P_{\cal R}$
is smaller than that in GR.
With the tensor power spectrum $P_{\rm T}$ obtained in Ref.~\cite{Cho:2014ija},
we observe that the tensor-to-scalar ratio in EiBI gravity becomes smaller than that ($r^{\rm GR} \sim 0.131$) in GR.
This reduction is affirmative in considering the dispute between the BICEP2 and the PLANCK results in the literature.
If a more precise value of $r$ is achieved from the observational results soon in the future,
it can provide a constraint on the value of $\kappa$ from our result.

\section*{Acknowledgement}
The authors are grateful to Hyeong-Chan Kim, Jinn-Ouk Gong and Macarena Lagos for useful discussions,
and to the Asia Pacific Center for Theoretical Physics (APCTP) for the hospitality.
This work was supported by the grant from the National Research Foundation
funded by the Korean government, No. NRF-2012R1A1A2006136 (I.C.).

\end{document}